\documentclass[12pt]{iopart}
\usepackage{iopams}  
\usepackage{multirow}
\usepackage{graphicx}
\usepackage{epsfig}
\begin{document}

\title{Highlights from PHENIX - I}

\author{A Franz (for the PHENIX Collaboration
\footnote{A list of members of the PHENIX Collaboration can be found at the end of this issue})}
\address{Brookhaven National Laboratory, Upton, NY 11973-5000 USA}
\ead{achim@bnl.gov}
\begin{abstract}

This contribution highlights recent results from the PHENIX Collaboration at RHIC. It covers global variables, flow and 2--particle correlations. A second contribution in this issue, by T.C.Awes, covers PHENIX results on heavy quarks, leptons and photons.

\end{abstract}

\submitto{\JPG}

\section{Introduction}

The Relativistic Heavy Ion Collider (RHIC) at the Brookhaven National Laboratory (BNL) in Upton, NY has just finished its 8$^{th}$ year of operation. The PHENIX Collaboration with its to--date 476 scientists from 67 institutions and 14 Nations has collected in the recent d--Au and p--p run a record 577 TB of data and 275 billions events. The Run--8 d--Au sample represents a 30 times increase over the Run--3 data set despite the addition of new detectors which are described in Sect.~\ref{sec:newdet} in more detail. These large data samples allow us to probe the properties of the new matter with precision measurements of the distributions and systematic study of their dependence on colliding system, centrality, rapidity or even the reaction plane. RHIC also increased its luminosity by a better understanding of the machine and new techniques like stochastic cooling.

RHIC is likely the most versatile heavy ion collider in the world and has collided in its first 8 years 4 different species at 6 different beam energies. 
Table~\ref{tab:runs} shows a summary of these first 8 years of PHENIX data taking.

\begin{table}[h]
\caption{Summary of the first 8 years of RHIC running for the PHENIX experiment}
\begin{center}
\begin{tabular}{|c|c|c|r|r|r|r|}\hline
\multirow{2}{*}{} &\multirow{2}{*}{Year}     &\multirow{2}{*}{Species}  & \multirow{2}{*}{$\sqrt{s}$ [GeV ]} & \multirow{2}{*}{$\int$Ldt}& N$_{tot}$&\multirow{2}{*}{Data Size} \\ 
                             &              & &  &  & (sampled) & \\ \hline
Run--1     & 2000    &Au--Au   &  130 &  1 $\mu$b$^{-1}$       & 10 M       &    3 TB     \\ \hline
\multirow{3}{*}
{Run--2} & \multirow{3}{*} {2001/02} & Au--Au &  200 &  24  $\mu$b$^{-1}$    & 170 M    &   10 TB    \\ \cline{3-7}
                &                 & Au--Au &  19    &                                       & $<$ 1 M &                  \\ \cline{3-7}
                &                 & p--p      &  200  &  0.15 pb$^{-1}$          & 3.7 B       &   20 TB    \\ \hline
\multirow{2}{*}
{Run--3} & \multirow{2}{*}{2002/03} & d--Au   &  200  &  2.74 nb$^{-1}$          & 5.5 B       &   46 TB    \\ \cline{3-7}
                &                 & p--p      &  200  &  0.35 pb$^{-1}$          & 6.6 B       &   35 TB    \\ \hline
\multirow{2}{*}
{Run--4} & \multirow{2}{*} {2003/04} & Au--Au &  200  &  241  $\mu$b$^{-1}$ & 1.5 B       &  270 TB   \\ \cline{3-7}
                &                 & Au--Au &  62.4  &  9  $\mu$b$^{-1}$     & 58 M       &   10 TB     \\ \hline
\multirow{4}{*}
{Run--5} & \multirow{4}{*}{2005}      & Cu--Cu &  200   &  3  nb$^{-1}$              & 8.6 B      &  173 TB    \\ \cline{3-7}
                &                 & Cu--Cu &  62.4  &  0.19 nb$^{-1}$         & 0.4  B     &   48 TB     \\ \cline{3-7}
                &                 & Cu--Cu &  22.4  &  2.7 $\mu$b$^{-1}$   & 9 M        &    1 TB       \\ \cline{3-7}
                &                 & p--p       &  200   &  3.8 pb$^{-1}$            & 85 B      &  262 TB     \\ \hline
\multirow{2}{*}
{Run--6} & \multirow{2}{*}{2006}       & p--p      &  200    &  10.7 pb$^{-1}$          & 233 B   &  310 TB     \\ \cline{3-7}
                &                  & p--p      &  62.4   &  0.1 pb$^{-1}$            &  28 B    &   25 TB       \\ \hline
Run--7   & 2007        & Au--Au &  200    &  813 $\mu$b$^{-1}$ &   5.1 B  &  650 TB     \\ \hline
\multirow{3}{*}
{Run--8} & \multirow{3}{*}{2007/08}  & d--Au   &  200     &  80 nb$^{-1}$             & 160 B  &  437 TB     \\ \cline{3-7}
                &                  & p--p      &  200     &  5.2 pb$^{-1}$            & 115 B  &  118 TB     \\ \cline{3-7}
                &                  & Au--Au &  9.2      &                                       &             &  few k         \\ \hline
\end{tabular}
\end{center}
\label{tab:runs}
\end{table}

The different collisions systems varying from simple p--p and d--Au, where cold nuclear effects should be visible, serving as a baseline and via proper scaling as a comparison to the collisions of heavier ions, e.g. Cu--Cu and Au--Au. These comparisons should enhance the difference of scaled p--p collisions to the properties of the produced dense medium. In 2003 all four RHIC experiments published white papers \cite{whitepaper} to summarize their findings which led to the announcement that a new phase of matter had been found \cite{perfectliquid}.

\section{New PHENIX detector subsystems} \label{sec:newdet}

\begin{figure}[h]
	\begin{minipage}[b]{0.45\linewidth}
		\centering
		\includegraphics[width=\linewidth]{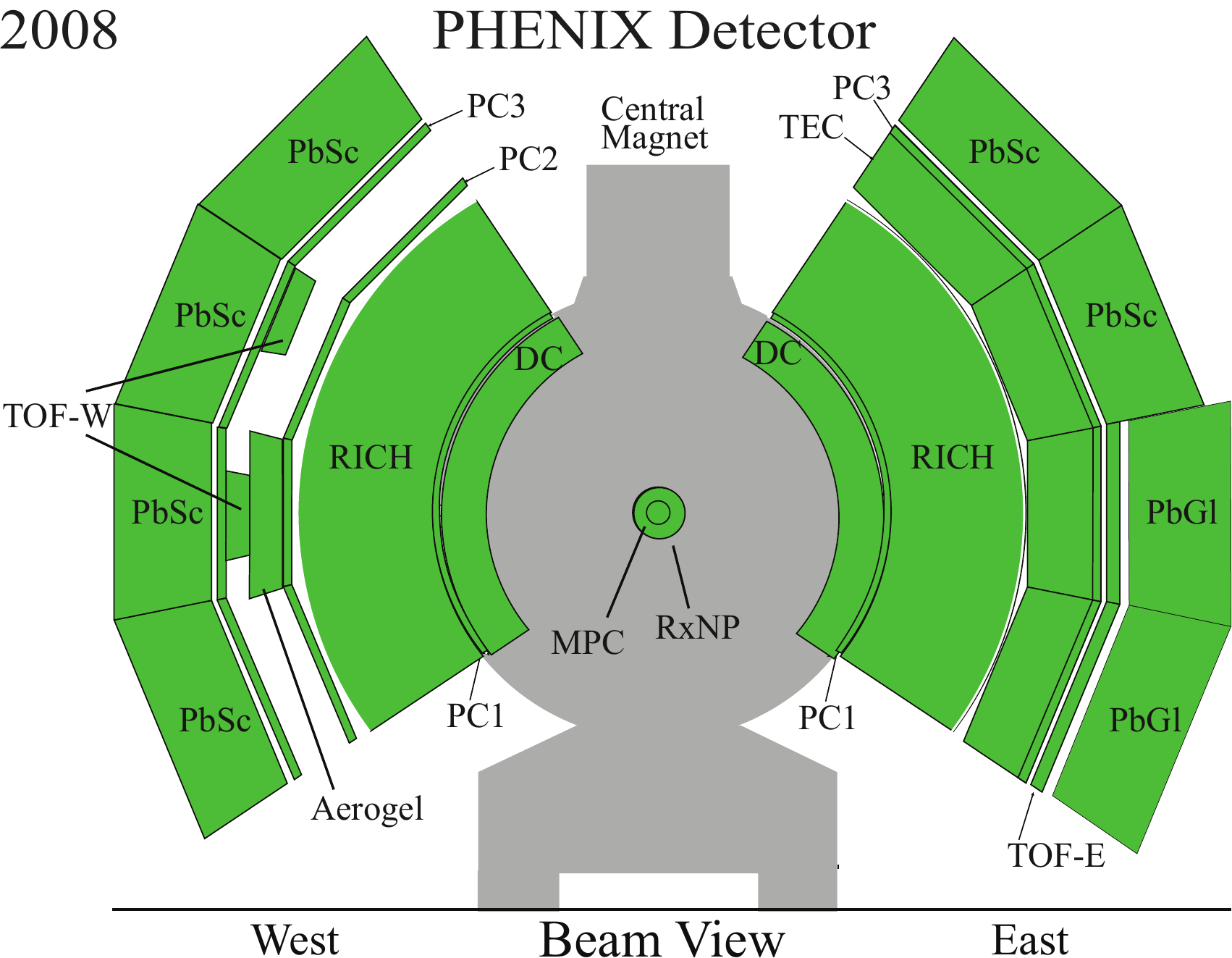}
		\caption{{\bf View of the PHENIX detector in beam direction.}}
		\label{fig:beamview}
	\end{minipage}%
	\begin{minipage}[b]{0.55\linewidth}
		\centering
		\includegraphics[width=\linewidth]{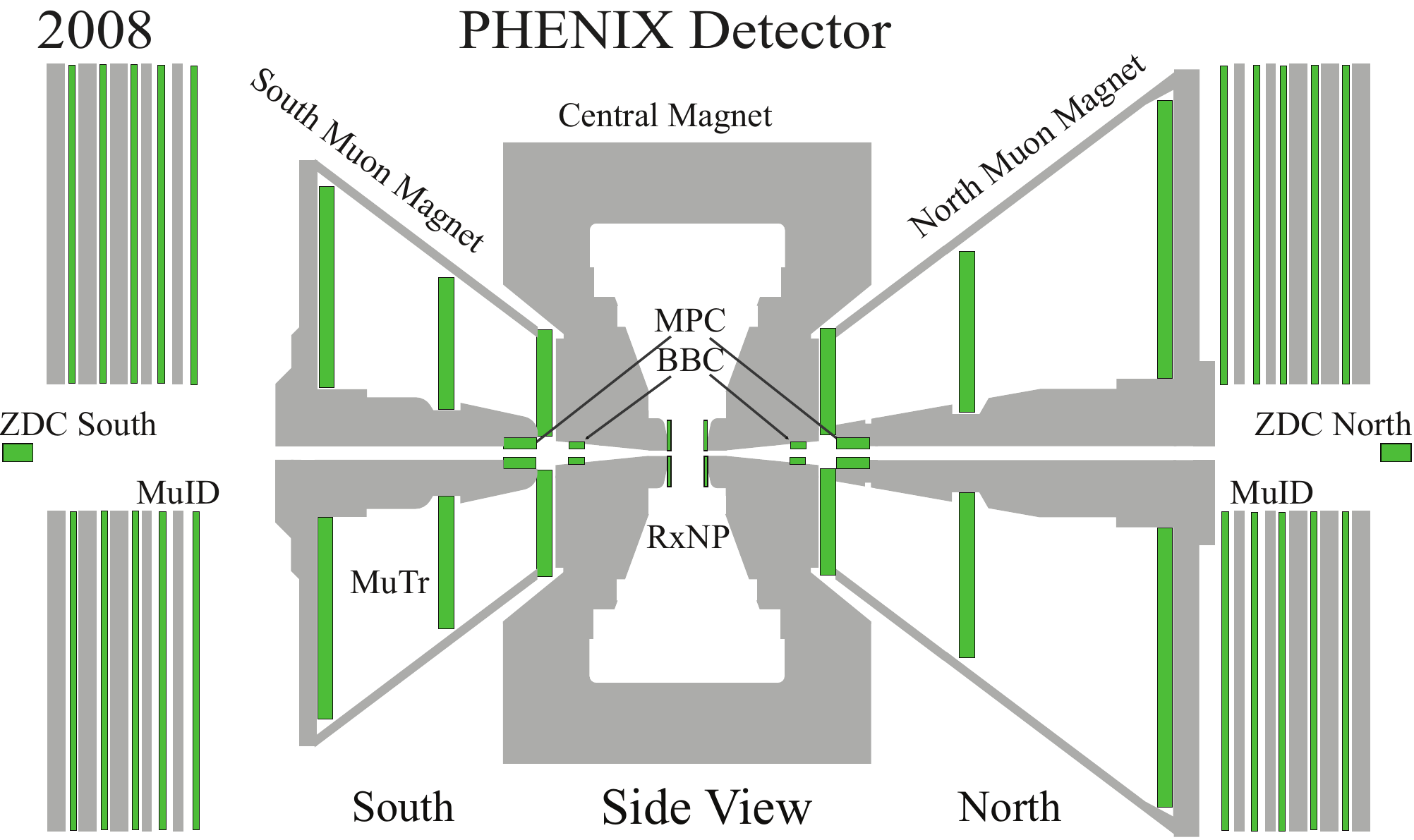}
		\caption{{\bf  View of the PHENIX detector in side--view.}}
		\label{fig:sideview}
	\end{minipage}
\end{figure}

\Fref{fig:beamview} and \ref{fig:sideview} 
show the PHENIX detector in the 2007/2008 configuration. It consists of 4 spectrometer arms with 3 main magnets. Two arms at mid-rapidity (East and West) with tracking, particle identification (PID) detectors and calorimeters for hadron, electron and photon detection and two muon arms in the forward angles (North and South). Details can be found in \cite{phenixnim}.

Several new detectors had been added over the past years to improve PID, the measurement of the reaction plane (RP) and $\pi^{0}$ identification at forward angles.

A time-of-flight (TOF-W) detector,  based on multi-gap resistive plate chambers (MRPC) \cite{tofw}, was added to the PHENIX West central arm detector in 2007 to extend the PID to higher momenta, i.e. above 2-3 GeV/c. Before the PID in the PHENIX West arm relied on a combination of a gas Ring Image Cherenkov (RICH) vessel, an aerogel detector (n=1.0114), and the electromagnetic calorimeter (EMCal) which  left a gap in the pion to kaon separation between 3 and 5 GeV/c. One octant of these MRPCs with pad readout were installed and achieving a 75ps time resolution, 85ps overall with the Beam--Beam Counter (BBC), the interaction and TOF start detector, resolution folded in. Several new results based on this detector have been presented at this conference \cite{Shengli,Valle}.

For a few years PHENIX will use the new Reaction Plane Detector (RxNP) \cite{rxnp} to further improve the RP measurement and to improve triggering at lower energies when the BBC and Zero--Degree--Calorimeters (ZDC) are not efficient enough. The RxNP consists of 2x2 rings with 12 scintillator 
counters each, read out by 2x24 photomultipliers. It covers the pseudorapidity windows $\eta=1.0\rightarrow1.5, 1.5\rightarrow2.8$. 

Seeking to extend the detector coverage into the forward direction, PHENIX installed 412~PbWO$_{4}$ crystals into the forward tips, $3.1<|\eta|<3.7$, of each magnet piston in the North and South muon magnets. The main goal for the Muon Piston Calorimeter (MPC) \cite{mpc}, as it is called, is the reconstruction of $\pi^{0}$  and the search for spin asymmetries in p--p collisions. In heavy ion running, when the overall multiplicity is too high, it improves the measurement of the event reaction plane.

A detector which had a first engineering run in 2007 is the Hadron Blind Detector (HBD) \cite{hbd1,hbd2}. Its a windowless Cherenkov detector using pure $CF_{4}$ with a triple GEM readout, where the top most layer is coated with Cesium Iodide (CsI) to convert the Cherenkov photons into photo-electrons which are in turn amplified by the GEM with a gain of $\sim5\cdot10^{3}$. The HBD will be important in coming years for the measurement of low mass electron pairs from the decay of light vector mesons ($\rho$, $\omega$, and $\phi$).

In the coming years PHENIX plans to install a silicon vertex tracking system, see \cite{Hubert}, a forward silicon--tungsten calorimeter, and a muon trigger based on resistive plate chambers.

\section{Global Observables}\label{sec:global}
Presentations at Quark Matter 1987 in Nordkirchen \cite{nordkirchen}, more than 20 years ago, concentrated on the first measurements of global observables like event multiplicity, transverse momentum and energy distributions to understand if the levels of energy densities reached where sufficient to form a QGP. More than ten years later, in 2000, further detailed measurements lead to the announcement \cite{cern} that a new state of matter had been observed. 

\begin{figure}[h]
	\begin{minipage}[b]{0.5\linewidth}
		\centering
		\includegraphics[width=\linewidth]{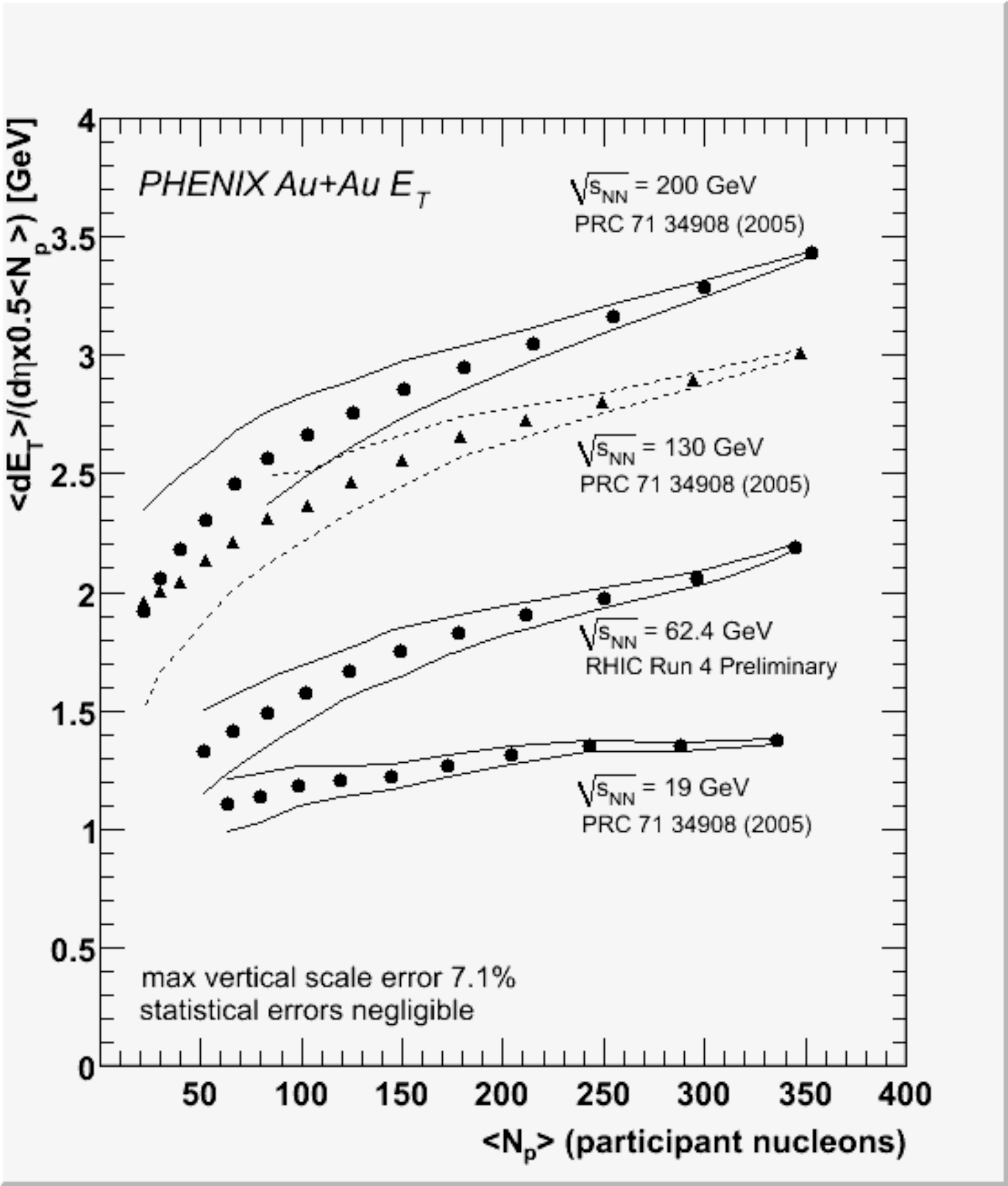}
		\caption{{\bf Transverse energy versus number of participating nucleons in Au~-~Au collisions.}}
		\label{fig:etauau}
	\end{minipage}%
	\begin{minipage}[b]{0.5\linewidth}
		\centering
		\includegraphics[width=\linewidth]{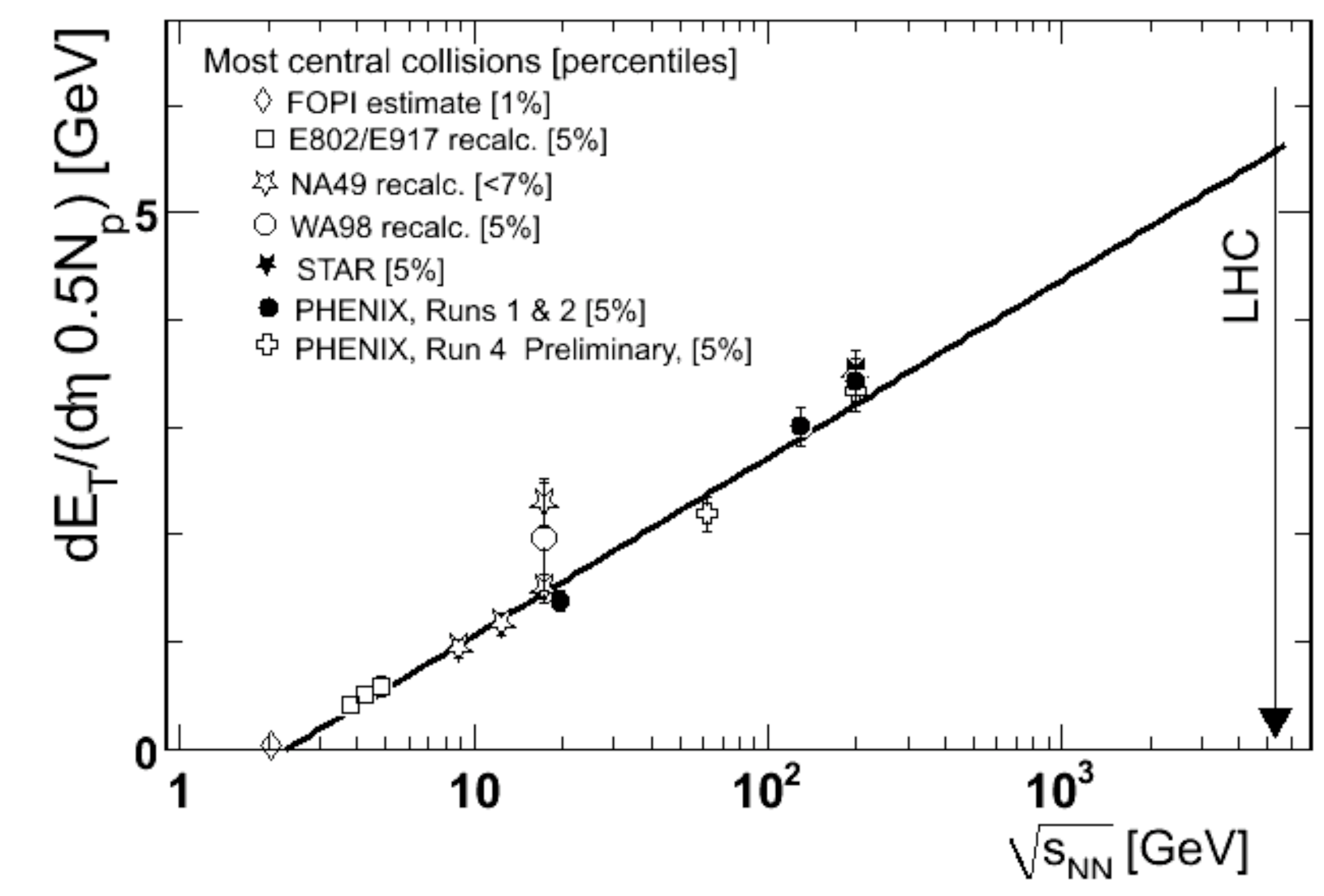}
		\caption{{\bf Transverse energy versus $\sqrt{s}$ for central Au~-~Au collisions.}}
		\label{fig:etcent}
	\end{minipage}
\end{figure}

PHENIX had published measurements of the total transverse energy, $E_{T}$, for $\sqrt{s}$~=~200, 130 and 13.9~GeV/c previously \cite{et1, et2}. 
\Fref{fig:etauau} summarizes these measurements for Au--Au collisions as a function of participating nucleons and adds the distribution for the fourth beam energy. $E_{T}$ increases with increasing number of participants and stronger with higher beam energy. Concentrating on the most central collisions, 
\Fref{fig:etcent} shows $dE_{T}/(d\eta\ 0.5N_{p})$ as a function of $\sqrt{s}$ for several measurements including the new PHENIX datapoint at $\sqrt{s}$~=~62.4~GeV/c. The measurement falls well in line with the previous observed linear dependence of the scaled $E_{T}$ with the log of $\sqrt{s}$.

Detailed studies of charged particle multiplicities in smaller and smaller rapidity windows are accessible with the large datasets. In this volume PHENIX presents fluctuation studies in overall multiplicities and particle ratios \cite{Homma}. Deviations from a monotonic behavior in these ratios should indicate a possible phase transition or critical change in the medium. In the current data nothing unusual was found.

A long proven approach to study the dynamics of a heavy ion collisions is a Hanbury-Brown and Twiss, HBT, analysis \cite{hbt1, hbt2}. PHENIX presented first results in reconstructing the shape and dynamical evolution of the particle emitting source by a 3--dimensional source imaging technique \cite{Lacey}.

Another method to estimate the source size is to measure the coalescence parameter, $B_2$, for deuterons \cite{deuterons, Valle}. Using the above mentioned TOF-W detector PHENIX could extend the existing (anti--) deuteron measurements to higher $p_T$ and multiple centrality bins. Expressing the coalescence probability, $B_{2}^{-1}$ is a measure of the source radius. It increases linearly with the number of participating nucleons in the collision, and the extracted radius parameter is compatible with HBT results on pion pairs.

\section{Flow}
A most surprising observation at RHIC was the strong elliptic flow, which lead to the conclusion that the medium we are studying does not behave like a hot gas but rather like a strongly coupled liquid. When colliding at intermediate impact parameters  the overlap region between the two nuclei is elliptically shaped in the transverse plane. This spacial anisotropy creates a pressure gradient which translates into a momentum anisotropy in the final particle stage. Experimentally this is measured via the $\phi$ angular distribution of the particles with respect to the reaction plane angle, $\Psi_{R}$, of the event which is defined by the beam direction and the distance vector of the center of the two nuclei \cite{flowintro}, and a Fourier decomposition.
\begin{equation}
E\frac{d^{3}N}{dp^{3}}=\frac{1}{2\pi }\frac{dN}{p_{T}dp_{T}dy}\left[%
1+\sum_{n=1}^{\infty }2v_{n}(p_{T},y)\cos (n\phi )\right]  \label{dndphi}
\end{equation}%
Because of the symmetry $\phi \leftrightarrow -\phi $ in the collision
geometry, sine terms do not appear in above expansion. Also the
odd-order anisotropic flows of particles at midrapidity vanish in collisions
with equal mass nuclei as a result of the additional symmetry $\phi
\leftrightarrow\phi +\pi $.

\begin{figure}[h]
	\begin{minipage}[b]{\linewidth}
		\centering
		\includegraphics[width=\linewidth]{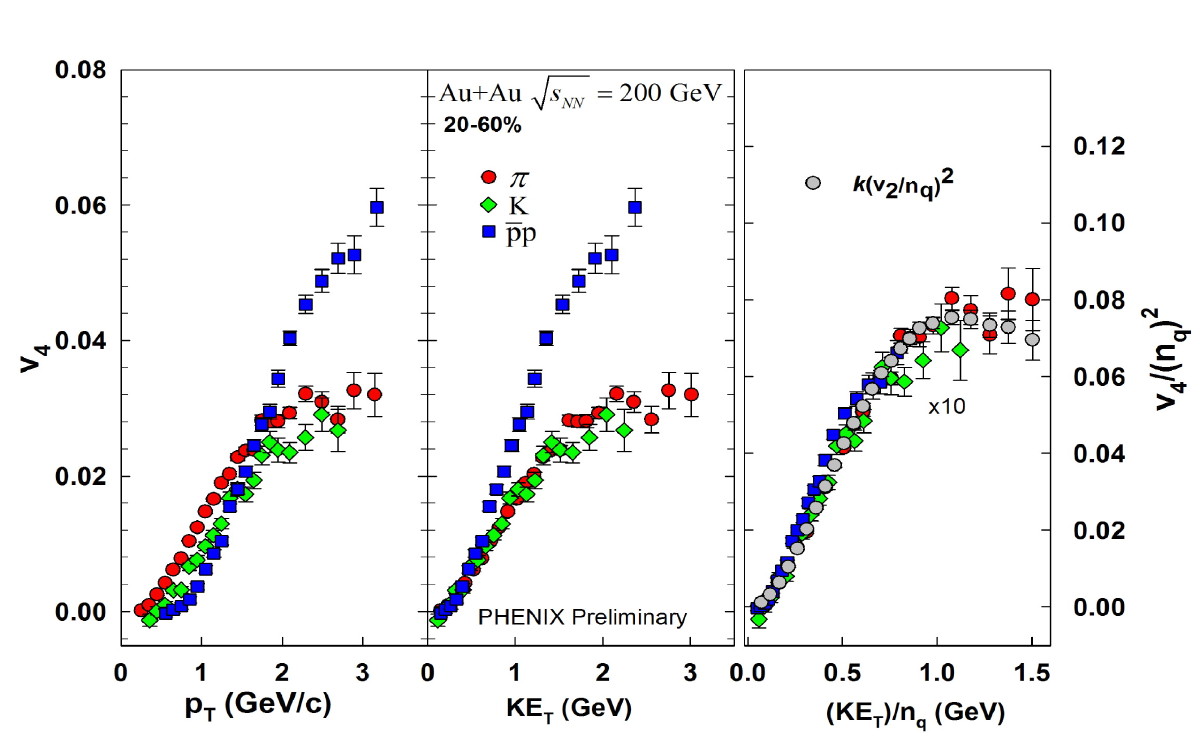}
		\caption{{\bf $v_4$ as a function of $p_T$, $KE_T$ and $KE_{T}/n_{q}$}}
		\label{fig:v4ket}
	\end{minipage}
\end{figure}

The second coefficient in the Fourier transform, $v_{2}$, is usually the largest and has been studied by all RHIC experiments. It has been observed that the $v_{2}$ of all studied particles scales as $v_{2}/n_{q} \sim KE_{T}/n_{q}$, where $n_{q}$ is the number of quarks in the particle and $KE_{T}=m_{T}-m_{0}$ is the transverse kinetic energy. This scaling was observed up to $KE_{T}\approx1~GeV/c$, an indication that hydrodynamical description of the data was valid \cite{Taranenko}. New data from PHENIX \cite{Shengli} indicate that this is not valid above 1~GeV/c which corresponds to $p_{T} \approx 3~GeV/c$ for a proton, the region where hard scattering becomes important.  

The next higher term $v_{4}$ is an important measure if a ideal hydrodynamical description is applicable in this momentum range. If valid, $v_{4}$ should follow the same scaling in $KE_{T}$ as $v_{2}$, but scaled with $n_{q}^{2}$ and more important be equivalent to $v_{2}^{2} n_{q}^{-2}$ as demonstrated in the right panel of \Fref{fig:v4ket}.

\begin{figure}[h]
	\begin{minipage}[b]{\linewidth}
		\centering
		\includegraphics[width=\linewidth]{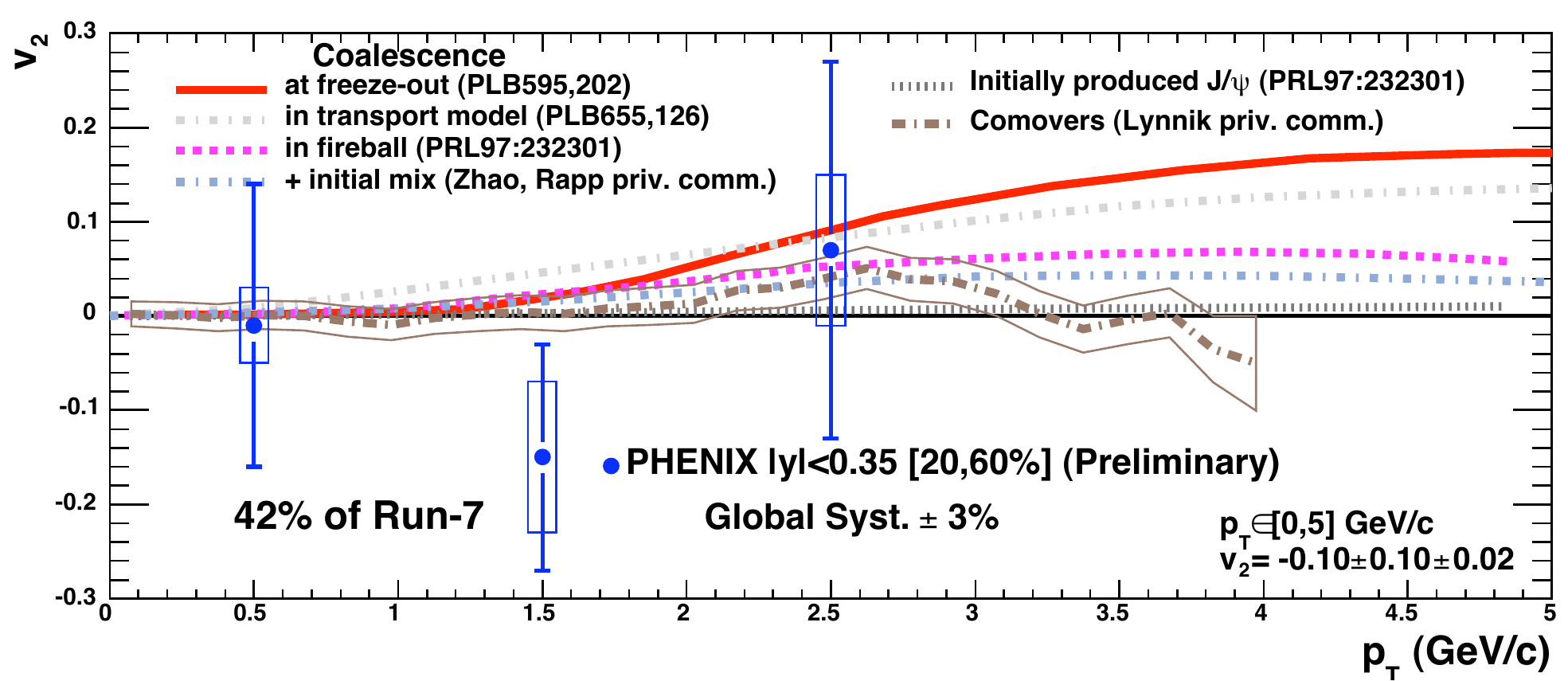}
		\caption{{\bf $v_2$ for $J/\Psi \rightarrow e^+e^-$} measured at mid-radidity in Au -- Au collisions with the PHENIX central arm detectors. 
		The lines indicate theoretical predictions as indicated in the figure.}
		\label{fig:v2jpsi}
	\end{minipage}
\end{figure}

\Fref{fig:v2jpsi} shows the first data on the $v_2$ of $J/\Psi \rightarrow e^+e^-$ in heavy ion collisions. The preliminary result of $v_2=-0.10\pm0.10\pm0.02$ is compatible with 0, but only 42\% of the data are analyzed so far. The lines in \Fref{fig:v2jpsi} represent theoretical predictions as indicated. A PHENIX result on $J/\Psi \rightarrow \mu^+\mu^-$ is to be presented soon, see \cite{Sylvestre} for more details.

Instead of a Fourier transform of the angular particle distribution with respect to the reaction plane, a cumulant technique \cite{cumulants, Issah} can also be used to extract the anisotropic flow strength. The detectors used for the RP determination in PHENIX have a large enough rapidity gap to the tracking and PID detectors that non-flow effects are not important, A direct comparison should then indicate where non--flow effects set in. The cumulant $v_{2}$ starts to diverge from the RP $v_2$  at $p_{T} \approx 3.5GeV/c$, indicating that non--flow effects, e.g. jets from hard scattering, become important. Incidentally it is also the region where the $KE_{T}/n_{q}$ scaling starts to fail. 

\section{Jets}

PHENIX has studied the properties of jets at 200 and 62.4~GeV/c for p--p, d--Au, Cu--Cu, and Au--Au collisions \cite{jets0, jets1, jets2, jets3, Adare, McCumber, Pei}. 

\begin{figure}[h]
	\begin{minipage}[b]{\linewidth}
		\centering
		\includegraphics[width=0.6\linewidth]{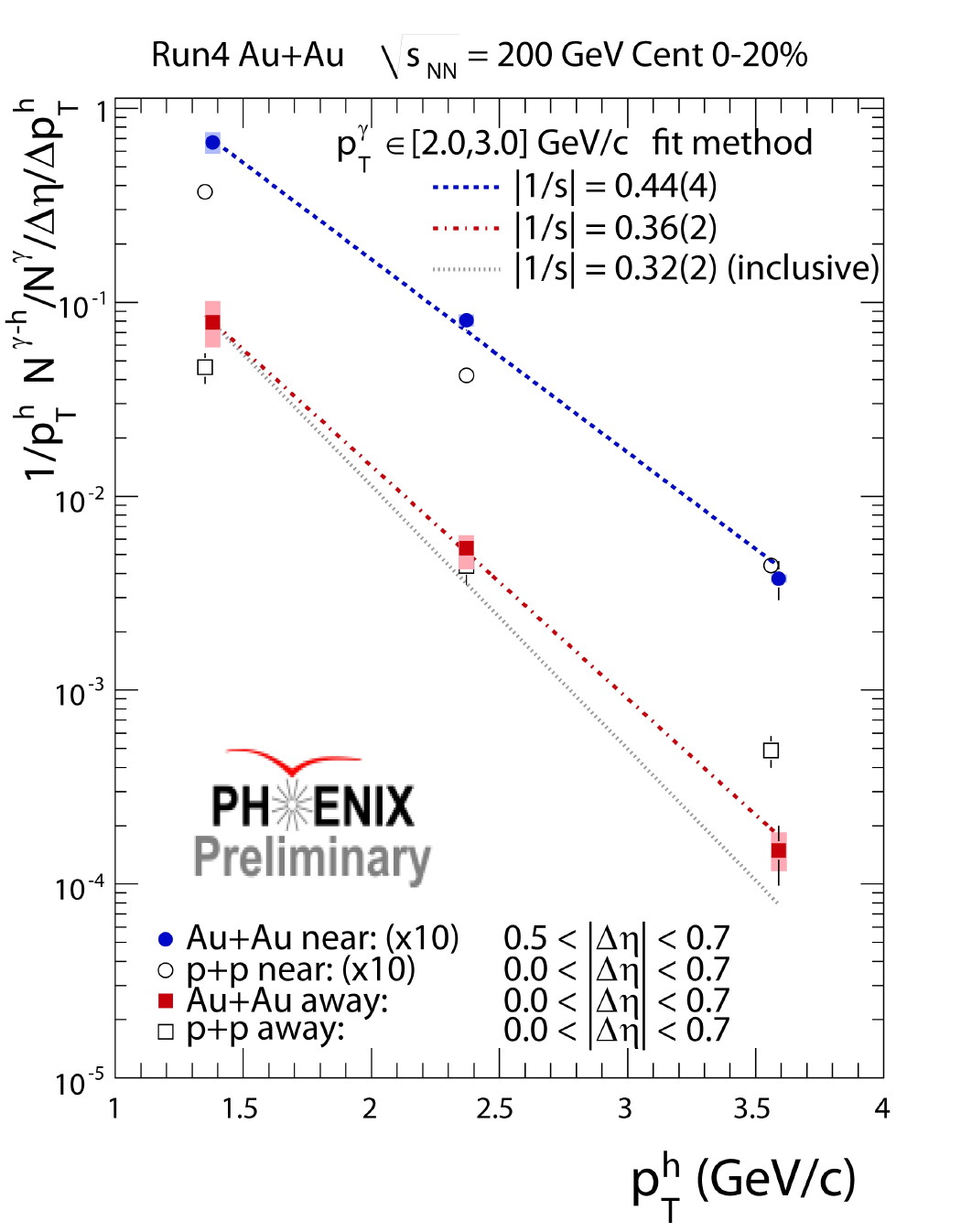}
		\caption{{\bf Near (squares) and awayside (circles) transverse momentum spectra for Au--Au (filled) and p--p (open) collisions for 
		different rapidity regions for the near and away side jets.
		The lines represent fits to the datapoints, with the solid line indicating the fit to an inclusive $p_T$ distribution.}}
		\label{fig:nearaway}
	\end{minipage}
\end{figure}

Jets resulting from a hard scattering of partons are impossible to reconstruct in heavy ion collisions because of the high background. Therefore these studies are done by 2 and 2+1 azimuthal correlations of a high $p_T$ particle, assumed to be the leading particle of one jet--arm and all other particles assumed to be from the same jet or the recoil. The correlation functions have to be corrected for background and flow, which in itself is an angular correlation. 

It has been observed in p--p and peripheral A--A collisions that opposite the trigger particle jet (near side) a slightly wide correlated distribution emerges (away side). In central A--A collisions the momentum spectra for the away side softens and the angular distribution widens even more. Several explanations for these effects have been presented at this conference. \Fref{fig:nearaway} shows a PHENIX comparison of the momentum distribution for two pseudo-rapidity, $\eta$, regions of the near and away side in p--p and Au--Au collisions. The momentum spectra for Au--Au collisions on the near side, but away from the main jet (upper, blue, solid points), and the away side (lower, red, solid points) are softer compared to p--p (open points) and close to the inclusive spectra (lines). This indicates that the momentum distributions of these particles has been softened by passing through the medium.

\begin{figure}[ht]
	\begin{minipage}[b]{\linewidth}
		\centering
		\includegraphics[width=\linewidth]{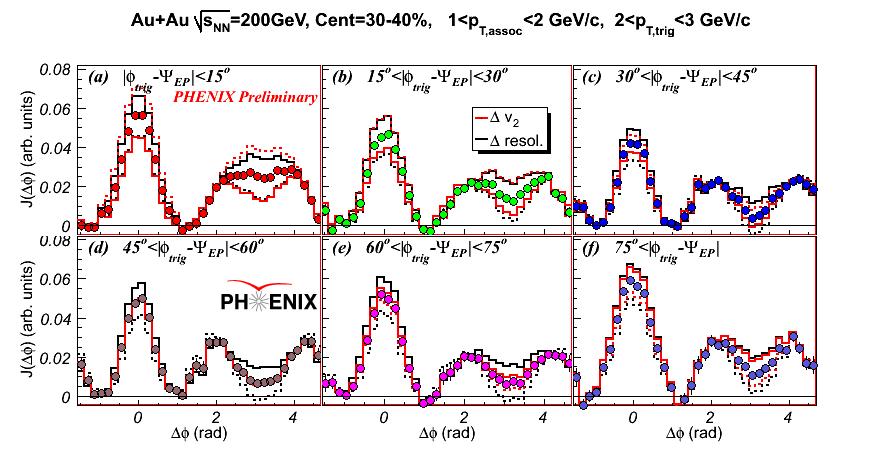}
		\caption{{\bf Jet correlations functions for Au Au collisions, panels represent 15 degree slices from 0 - 90 degrees away from the reaction plane.}}
		\label{fig:rpcorrelations}
	\end{minipage}
\end{figure}

If the particle distributions and momenta are affected by their passage through the collision medium than the distributions when measured along the long versus the short axis of the collisions ellipsoid should be different. \Fref{fig:rpcorrelations} shows preliminary PHENIX results on a 2 particle correlation function were the data are binned in angular regions with respect to the reaction plane. Even so the $v_2$ dominated systematic error is large a clear change in the shape of the distributions is visible.

\section{Summary}

PHENIX has collected a vast sample of data from p--p to Au--Au collisions at various energies. The data show that we have created a dense medium which affects the momenta and angular distributions of the produced particles. On the other hand it shows a strongly coupled flow which affects all produced particles, even heavy quarks. PHENIX has shown multiple new and more detailed results at this conference and will with its current and future detector subsystems continue to uncover the details of this 'perfect liquid'.

\end{document}